\documentclass[11pt,twoside]{article}

\usepackage{asp2004}
\usepackage{epsf}
\usepackage{psfig}
\usepackage{lscape}

\begin{document}
\title{MIPS 70$\mu$m Observations of the Spitzer Extragalactic First Look Survey}   %%% Fill in title
\author{D. T. Frayer}   %%% Fill in author names
\affil{Spitzer Science Center, Caltech, 220-06, Pasadena, CA  91125}    %%% Fill in author affiliations

\begin{abstract} %%% Abstract to run on from here.

Early results from the 70$\mu$m observations of the Spitzer
Extragalactic First Look Survey are presented.  As a whole, about 90\%
of the population show infrared colors that are consistent with the
spectral energy distributions of the local IRAS population of luminous
infrared starbursts, while approximately 10\% of the sample show warm
infrared colors consistent with AGN activity.  The mean redshift for
the population is $z=0.2$, and the mean total infrared luminosity is
about $L({\rm ir})= 2\times 10^{11} L_{\odot}$.

\end{abstract}

%%% MAIN BODY OF TEXT GOES HERE. CONSULT "INSTRUCTIONS FOR AUTHORS USING
%%% LATEX2E MARKUP", SECTIONS 2.3-2.6 FOR HELP WITH EQUATIONS, FIGURES,
%%% AND TABLES.

%\vspace*{-1cm}
\section{Observations and Data Reduction}   

One of the first science observations carried out with the Spitzer
Space Telescope was the extragalactic First Look Survey (FLS) which
was designed to characterize the infrared sky at previously unexplored
sensitivities.  The extragalactic FLS covers a 4.4 sq-degree region
near the ecliptic pole (RA[J2000]=17:18, Dec[J2000]=+59:30), within
the northern continuous viewing zone of Spitzer.  Inside the FLS main
survey a smaller verification strip of 0.25 sq-degree was observed
with an integration time of 4 times that of the main survey to
characterize the completeness and source reliability of the main
survey.  In total, 27.7 hours of observations were taken in 2003
December with the Multiband Imaging Photometer for Spitzer (MIPS,
Rieke et al. 2004).

The basic calibrated data products (BCDs) used were downloaded from
the SSC data archive, software version S10.  The general steps for
processing MIPS data are described in Gordon et al. (2005).  The two
main artifacts impacting the FLS-70$\mu$m data are the stim flash
latents and the variations of the slow response as a function of time.
The combination of a high-pass time median filter and a column median
filter removes the bulk of the data artifacts.  To preserve
photometric results, the data were filtered in two passes.  In the
first filtering pass, we applied the filtering corrections and coadded
the data to determine the positions of sources.  The source positions
within the original BCDs were flagged and new filtering corrections
were calculated in the second pass, ignoring the pixels containing
sources.  The two pass reduction minimized the data artifacts while
preserving point-source calibration.  The data were coadded and
corrected for array distortions with the SSC MOPEX software.  Source
detection and photometry were done using the SSC APEX software
package.  The achieved 5$\sigma$ point source sensitivities are
20\,mJy for the main survey (42\,s of integration) and about 10\,mJy
for the verification strip (210\,s of integration).

\section{Results and Conclusions}

%\vspace*{-2mm}
\begin{figure}[!ht]
\plotfiddle{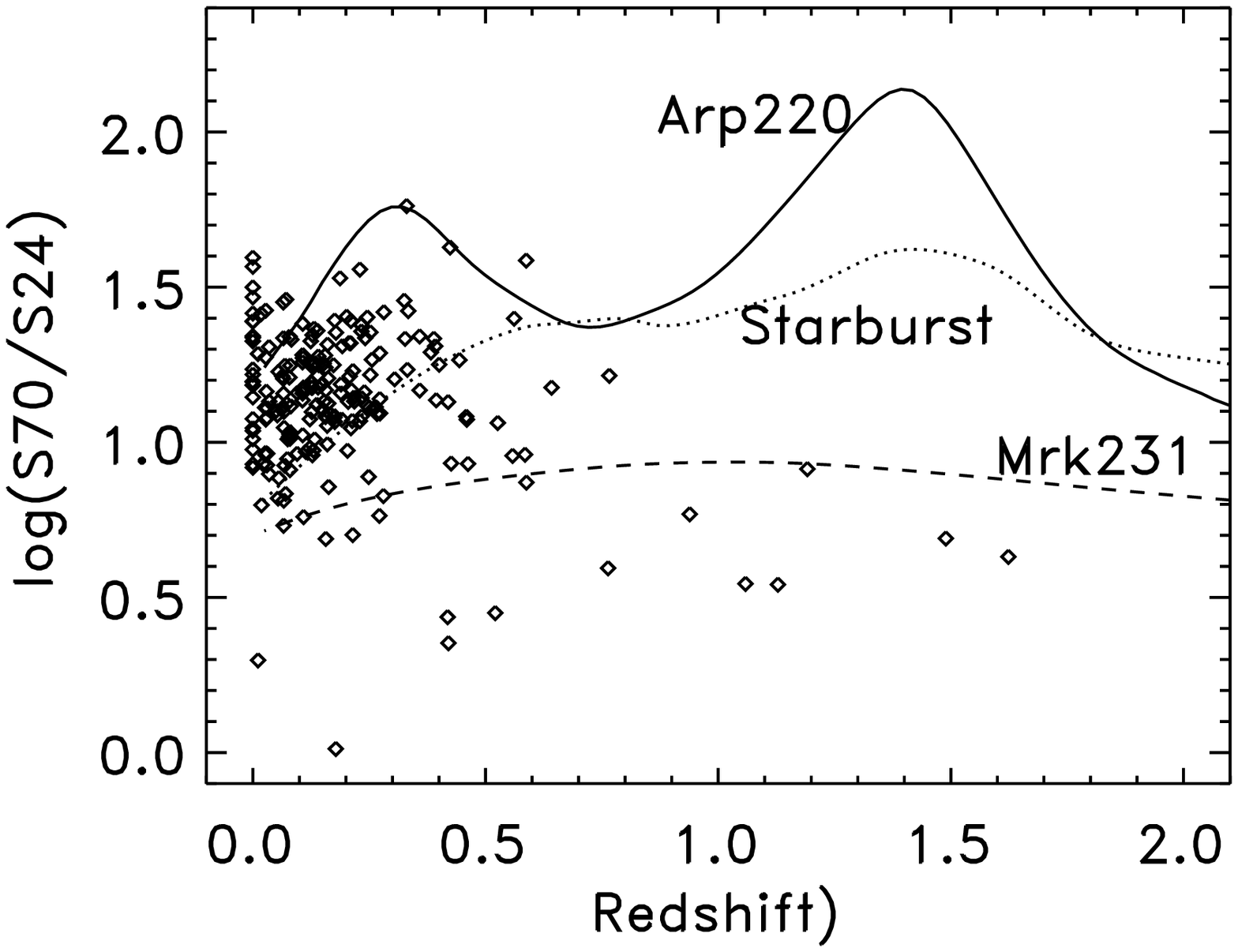}{4.cm}{0.}{40}{40}{-212}{-70}
\plotfiddle{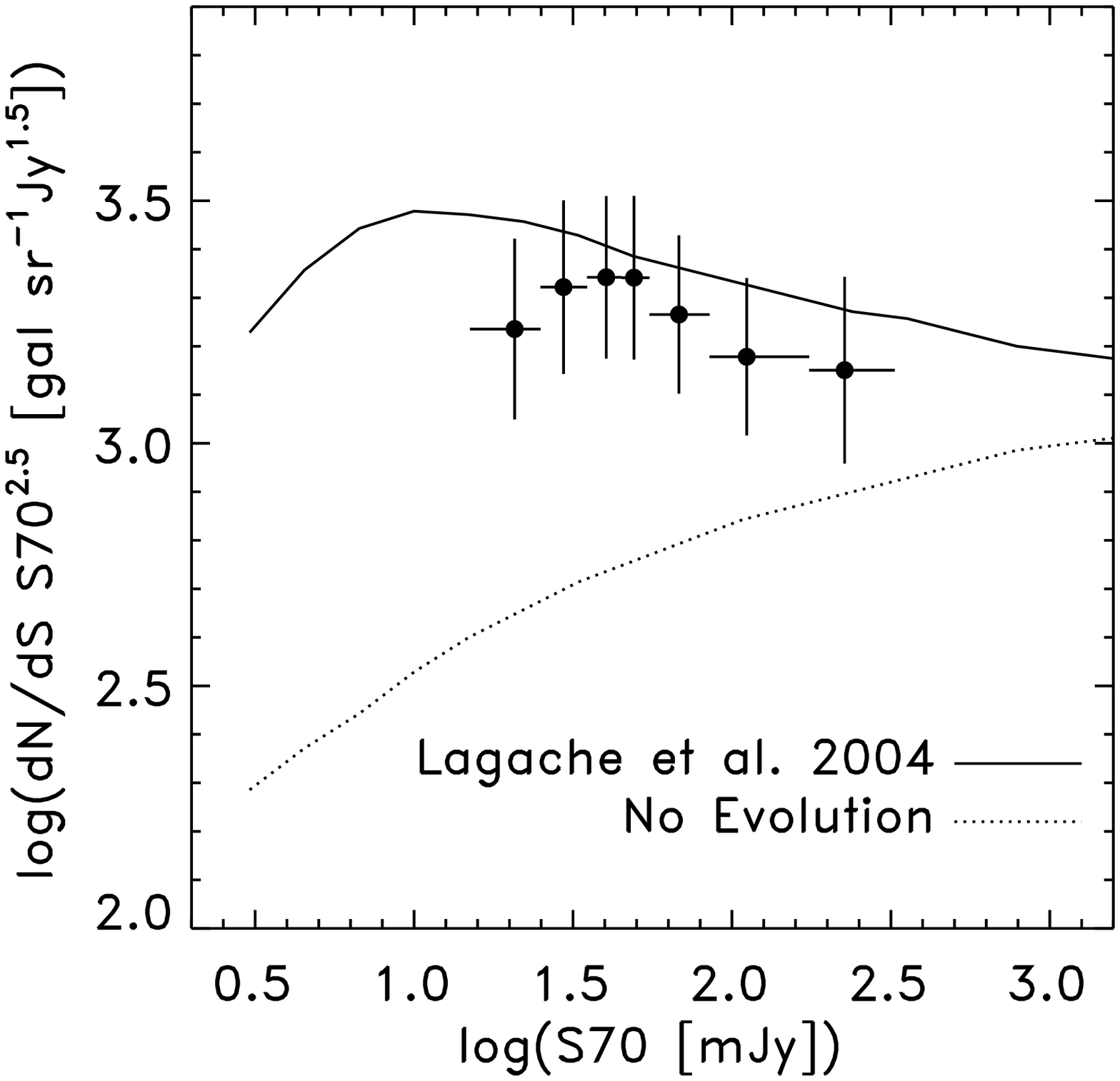}{0cm}{0.}{35}{35}{-1}{-60}
%\plottwo{frayerd_fig1.ps}{frayerd_fig2.ps}
\end{figure}

%\vspace*{-2mm}

{\bf Figure 1 (left)} shows the observed infrared colors (flux ratios)
for the 70$\mu$m-selected FLS sources with known redshifts compared to
the predicted colors as a function of redshift. The local ULIRGs
Arp220 (starburst) and Mrk231 (AGN+starburst) are shown by the solid
and dashed lines, respectively.  A starburst model from Dale \& Helou
(2002) is represented by the dotted line.  Only sources with
one-to-one band-merged matches are shown to avoid potential issues
with false matches.  The S70/S24 flux density ratios can be used to
help distinguish between infrared-warm AGN and infrared-cool
starbursts (S70/S24$\ga10$).  The majority of galaxies detected at
these depths are $z<1$ starbursts.  Sources below the dashed line of
Mrk\,231 are likely AGN.

{\bf Figure 2 (right)} shows the FLS 70$\mu$m differential number
counts (Frayer et al. 2005, in preparation) compared to the models of
Lagache et al. (2004).  The FLS 70$\mu$m counts are consistent with
the counts measured in other fields from Dole et al. (2004).

About 90\% of the sources with redshifts have infrared-cool starburst
colors, while about 10\% have warm colors consistent with AGN
activity.  For the 70$\mu$m selected-sources with known redshifts, the
mean redshift for the population is $z\simeq0.2$.  The average S70/S24
ratio is consistent with the ratio expected based on S60/S25 colors
observed for the IRAS population if redshifted to $z=0.2$.  Using the
models of Dale \& Helou (2002), we estimate a mean total infrared
luminosity of $L({\rm ir}) \simeq 2\times10^{11} L_{\odot}$ for these
sources.  Much deeper 70$\mu$m observations are needed to constrain
the evolutionary models at high-redshift and to estimate the
relative contribution of LIRGs and high-redshift ULIRGs to the total
cosmic infrared background.

%%% Top level section head (remove "%" symbol)
%\subsection{}   %%% Second level section head (remove "%" symbol)
%\subsubsection{}   %%% Lowest level section head (remove "%" symbol)
%\section*{}	%%% Unnumbered top level section head (remove "%" symbol)
%\subsection*{}   %%% Unnumbered second level section head (remove "%" symbol)

\acknowledgements %%% Text of acknowledgements runs on after this command.

I thank my SSC colleagues and the MIPS Instrument Team who have made
these observations and data reduction possible.

%, including Tom Soifer,
%George Helou, Lisa Storrie-Lombardi, Dave Shupe, Dario Fadda, Lin Yan,
%Phil Choi, Phil Appleton, Francine Marleau, David Henderson, David
%Makovoz, Ted Hesselroth, Misha Pesenson, Frank Masci, Mehrdad Moshir,
%Bill Latter, Alberto Noriega-Crespo, Stefanie Wachter, Debbie Padgett,
%Karl Gordon, Karl Misselt, Chad Engelbracht, John Stansberry, Doug
%Kelly, Jane Morrison, and George Rieke.
%%% THE BIBLIOGRAPHY
%%%
%%% CONSULT SECTION 3 OF "INSTRUCTIONS FOR AUTHORS" FOR HOW TO USE NATBIB.
%%% AUTHORS ARE ENCOURAGED TO USE EITHER THE "THEBIBLIOGRAPY" ENVIRONMENT
%%% BY UNCOMMENTING (DELETING THE "%" SYMBOL) THE COMMANDS BELOW, OR BY
%%% USING THE BIBTEX ENVIRONMENT. TO FIND OUT WHICH IS APPLICABLE TO YOUR
%%% CONTRIBUTION, CONSULT THE VOLUME EDITORS FOR YOUR PROCEEDINGS.
%%%

\end{document}